\documentclass[acmsmall,screen,nonacm]{acmart}

\usepackage{cleveref}
\usepackage[frozencache=true,cachedir=.]{minted}
\usepackage{caption}
\usepackage{subcaption}

\setminted{fontsize=\footnotesize}

\AtBeginDocument{%
  \providecommand\BibTeX{{%
    \normalfont B\kern-0.5em{\scshape i\kern-0.25em b}\kern-0.8em\TeX}}}

\begin{document}

\title{Design and Implementation of ShenWei Universal C/C++}

\author{Huanqi Cao}
\affiliation{
  \department{Department of Computer Science and Technology}
  \institution{Tsinghua University}
  \city{Beijing}
  \country{China}
}
\email{caohq18@mails.tsinghua.edu.cn}

\author{Jiajie Chen}
\affiliation{
  \department{Department of Computer Science and Technology}
  \institution{Tsinghua University}
  \city{Beijing}
  \country{China}
}
\email{cjj21@mails.tsinghua.edu.cn}



\begin{abstract}
The ShenWei many-core series processors powering multiple cutting-edge supercomputers are equipped with their unique on-chip heterogeneous architecture.
They have long required programmers to write separate codes for the control part on Management Processing Element (MPE) and accelerated part on Compute Processing Element (CPE), which is similar to open standards like OpenCL.
Such a programming model results in shattered code and bad maintainability, and also make it hard to migrate existing projects targeting commodity processors.
Borrowing the experience from CUDA and DPC++ and leveraging the unique unified main memory on ShenWei many-core architecture, we propose ShenWei Universal C/C++ (SWUC), a language extension to C/C++ that enables fluent programming acrossing the boundary of MPE and CPE.
Through the use of several new attributes and compiler directives, users are able to write codes running on MPE and CPE in a single file.
In case of C++, SWUC further support lambda expressions on CPEs, making it possible to have the code flow better matching the logical design.
SWUC also manages to make the Athread library interfaces available, easing the learning curve for original ShenWei users.
These powerful features together ensures SWUC to simplify the programming on ShenWei many-core processors and migration of existing C/C++ applications.
\end{abstract}

\maketitle

\section{Introduction}

The ShenWei many-core series processors, now containing SW26010 and SW26010pro, are adopted in Sunway supercomputers as their exclusive computing power source.
Sunway Taihulight~\cite{sunway} is powered by SW26010, and the Newest Generation Sunway Supercomputer~\cite{newsunway} is powered by SW26010pro.
Both chips share the unique on-chip heterogeneous architecture:
the Management Processing Elements (MPEs) are full-functional cores responsible for controlling, communication and file I/O, while the Compute Processing Elements are accelerator cores responsible for highly-efficient parallel computation.
They share a $1:64$ MPE to CPE counts ratio.

To program on ShenWei many-core series processors, users have to write and compile code separately for MPE and CPE.
The MPE part would be similar to conventional CPU codes, except that the accelerated part are replaced with kernel launches through the runtime libraries.
The CPE part, lying in other source files, implement the kernel functions launched by the MPE part.
The development workflow is thus similar to that of OpenCL~\cite{opencl}, except that the unified main memory makes things easier to implement the kernel.

Such a workflow introduces multiple issues in real application migration and development.
Besides the obvious issue that the code would be shattered to maintain, other problems can be more severe.
For example, migrating a hotspot loop in a well-abstracted application will require careful selection of \texttt{struct} definition and function implementation to be repeated on the CPE side.
Also, C++ applications with heavy template usage results in that we have to manually instantiate for every set of template parameters for a templated CPE kernel.

To address the above issues, we propose ShenWei Universal C/C++ (SWUC), which is a language extension to C/C++, inspired by CUDA~\cite{cuda} and OneAPI DPC++~\cite{dpc}.
SWUC do not alter the original hardware abstraction of ShenWei many-core architecture, but instead only aims at refining the programming experience.
It introduces four attributes, \texttt{host}, \texttt{slave}, \texttt{infer} and \texttt{kernel}, as well as a set of \texttt{\#pragma swuc} compiler directives to set the attributes in batch.
\texttt{host} and \texttt{slave} mark functions to be available on MPEs and CPEs, respectively.
The available sides of \texttt{infer} marked functions are determined automatically by functions called by them.
\texttt{kernel} functions are entry points of CPE threads, the invoke of which is translated into kernel spawns and represents launching the CPE array from MPE.

SWUC is then implemented as a plugin to the Sunway C/C++ compiler.
It runs a two-pass compilation to generate MPE and CPE machine code correspondingly.
To filter out code not available on current compiled side (MPE-only codes during CPE compilation or vice versa), it intercepts at the middle-end and removes code marked with the unavailable side.
The generated two objects are then linked together, simulating an end-to-end mixed compilation.

The above design and implementation make the SWUC a drop-in replacement for a conventional C/C++ compiler.
It enables straightforward compilation and smooth human-work curve during migration of applications.
The interception at middle-end and design of \texttt{infer} mode also greatly eases the use of templates across the boundary of MPE and CPE.
Altogether, SWUC opens a new era of C/C++ development on the ShenWei many-core architecture.

\section{Motivation}

In this section, we describe two typical scenario that the previous separated Management Processing Element (MPE) - Compute Processing Element (CPE) programming hurts productivity.

\begin{figure}[htbp]
    \centering

    \begin{subfigure}[t]{\linewidth}
    \begin{minted}[linenos]{cpp}
double helper(double a) { ... }
...
    for (int i = 0; i < N; ++i) {
        B[i] = helper(A[i]);
    }
...
    \end{minted}
    \subcaption{Original loop}
    \label{fig:helper-function:original}
    \vspace{1em}
    \end{subfigure}

    \begin{subfigure}[t]{\linewidth}
    \begin{minted}[linenos]{cpp}
double helper(double a) { ... }
...
    struct { double *A, *B; int N; } args = { A, B, N };
    athread_spawn(kernel, &args);
    athread_join();
...
    \end{minted}
    \subcaption{Accelerated: MPE code}
    \label{fig:helper-function:acc-mpe}
    \vspace{1em}
    \end{subfigure}

    \begin{subfigure}[t]{\linewidth}
    \begin{minted}[linenos]{cpp}
double helper(double a) { ... } /* copied to CPE code */
void kernel(struct { double *A, *B; int N; } *args) {
    double *A = args->A, *B = args->B;
    int N = args->N;
    for (int i = N * TID / 64; i < N * (TID + 1) / 64; ++i) {
        B[i] = helper(A[i]);
    }
}
    \end{minted}
    \subcaption{Accelerated: CPE code}
    \label{fig:helper-function:acc-cpe}
    \end{subfigure}

    \caption{Before and after migrating a parallelizable loop with a helper function to CPEs.}
    \label{fig:helper-function}
\end{figure}

\subsection{Migration of Existing Application}

To migrate an existing application targeting commodity clusters, especially CPUs, the first step would usually be identifying the parallelizable loops costing significant amount of time.
In a well abstracted application, it would be common to observe important \texttt{struct}s being referenced to and tool functions being called inside such a loop, like in \cref{fig:helper-function:original}.
In case of C++, it usually goes even further to member function calls.
After migration and parallelization, the MPE and CPE parts are listed in \cref{fig:helper-function:acc-mpe} and \cref{fig:helper-function:acc-cpe} respectively.
One has to first recursively identify the functions like the \texttt{helper} here, duplicate them in the CPE part, to make the loop run on CPEs, even if the \texttt{helper} itself might be a simple, pure and thread-safe function that should be directly available to the CPEs.
It can be imagined how sophisticated it would be to migrate a more complicated real world applicastion with C++, classes, member function calls and so on.

\subsection{Templated Kernels}

Templates in C++ are extremely helpful when we have generic codes work with different types and user-defined functions.
It has been a powerful basis for modern parallel programming libraries, including Thread Building Blocks~\cite{tbb}, Thrust~\cite{thrust} and so on.
However when used on ShenWei, we face the difficulty of collaborative template instantiation between MPE and CPE codes.

\begin{figure}[htbp]
    \centering

    \begin{subfigure}[t]{\linewidth}
    \begin{minted}[linenos]{c}
template <typename T>
void spawn__vector_add(std::tuple<T *, const T *, size_t> *args) {
    auto [a, b, n] = *args;
    constexpr size_t step = 4096 / sizeof(T);
    T la[step], lb[step];
    for (size_t i = step * CRTS_tid; i < n; i += step * 64) {
        size_t actual_step = std::min(step, n - i);
        dma_get(la, a + i, actual_step * sizeof(T));
        dma_get(lb, b + i, actual_step * sizeof(T));
        for (size_t ii = 0; ii < actual_step; ++ii)
            la[ii] += lb[ii];
        dma_put(a + i, la, actual_step * sizeof(T));
    }
}
    \end{minted}
    \subcaption{Templated CPE kernel function}
    \label{fig:template:kernel}
    \vspace{1em}
    \end{subfigure}

    \begin{subfigure}[t]{\linewidth}
    \begin{minted}[linenos]{c}
template <typename T>
void spawn__vector_add(std::tuple<T *, const T *, size_t> *args);
...
    std::tuple<int *, const int *, size_t> args{a, b, n};
    athread_spawn(spawn__vector_add<int>, &args);
...
    \end{minted}
    \subcaption{Invoking the templated kernel}
    \label{fig:template:invoke}
    \vspace{1em}
    \end{subfigure}

    \begin{subfigure}[t]{\linewidth}
    \begin{minted}[linenos]{c}
template void spawn__vector_add<int>(std::tuple<int *, const int *, size_t> *args);
template void spawn__vector_add<short>(std::tuple<short *, const short *, size_t> *args);
template void spawn__vector_add<long>(std::tuple<long *, const long *, size_t> *args);
template void spawn__vector_add<float>(std::tuple<float *, const float *, size_t> *args);
template void spawn__vector_add<double>(std::tuple<double *, const double *, size_t> *args);
template void spawn__vector_add<float16>(std::tuple<float16 *, const float16 *, size_t> *args);
    \end{minted}
    \subcaption{Manual instantiating the template on CPE side}
    \label{fig:template:instantiate}
    \end{subfigure}

    \begin{subfigure}[t]{\linewidth}
    \begin{minted}[linenos]{c}
struct int_float {
    int x;
    float y;
    int_float operator+(const int_float &other) const {
        return {x + other.x, y + other.y};
    }
};
template void spawn__vector_add<int_float>(std::tuple<int_float *, const int_float *, size_t> *args);
    \end{minted}
    \subcaption{Manual instantiation for user-defined type}
    \label{fig:template:instantiate-udt}
    \end{subfigure}

    \caption{Example for implementing, using and manually instantiating a templated CPE kernel function.}
    \label{fig:template}
\end{figure}

Consider the naive example shown in \cref{fig:template:kernel}.
It demonstrates a templated CPE kernel named \texttt{spawn\_\_vector\_add}, which implements adding two vectors for arbitrary type supporting the add operator, e.g. \texttt{int}, \texttt{float}, \texttt{double}, etc.
Then on MPE, we declare its prototype and use it like in \cref{fig:template:invoke}.
Both codes compile but the linking reports ``Undefined reference to \\ \texttt{slave\_\_Z17spawn\_\_vector\_addIiEvPSt5tupleIJPT\_PKS1\_mEE}''.
This is because that the CPE part is not aware of the instantiation of \texttt{spawn\_\_vector\_add<int>}, thus not instantiating it to a concrete function.
To resolve such issues, we have to add the manual instantiations in \cref{fig:template:instantiate}, each for one type parameter.
If some user plan to use the kernel with their own defined type, it would require instantiation like in \cref{fig:template:instantiate-udt}.
As the number of user-defined types and kernels growing, such manual instantiation quickly explodes to an unacceptable number.
For example, in the implementation of tensor operators in BaGuaLu~\cite{bagualu}, we wrote hundreds of such instantiation, which is painful.

Things get worse with the presence of user-defined functions.
In the implementation of Graph 500 on New Sunway \cite{graph500-swnew}, we implement the complicated message processing pipeline as templates to simplify the development.
However, such pipelines require types of user-defined functions (which is expected to be lambda expressions) as their template parameters to expand to efficient code.
The aforementioned template parallel libraries on commercial architectures also leverages user-defined functions to provide customizable parallel algorithms.
Under such circumstances on ShenWei, given the above analysis, one must express the types of lambdas to be able to write the manual instantiation, which is impossible.
Thus, one have to rewrite the lambda expressions to structs with \texttt{operator()}, greatly sacrificing the programmability.

\subsection{Design Goals of SWUC}

Based on the above examples representing our real world experiences, we list our design goals as below:

\begin{itemize}
    \item Unified single-source programming for code across MPE and CPE;
    \item Shared definition of \texttt{struct}s and \texttt{class}es for MPE and CPE;
    \item Collaborative instantiation of template functions between MPE and CPE;
    \item Complete support for lambda expressions across MPE and CPE.
\end{itemize}

Specifically, modifying the hardware abstraction is not our goal.
The previous hardware abstraction is widely accepted and we do not expect to change that.

\section{Language Extension Design}

In this section we introduce the design of SWUC as a language extension.

\subsection{Extension Attributes}

SWUC uses extension attributes to mark functions for different purposes.
It includes \texttt{host} and \texttt{slave} for marking available target (MPE and CPE correspondingly), \texttt{infer} for automatically inference of available target, and \texttt{kernel} for CPE kernel entry point.

\subsubsection{\texttt{host} and \texttt{slave} attributes}
These attributes simply mark a function to be available on that target.
By default a function is assumed as \texttt{host}.
They can be applied simultaneously to one function, indicating that the function can be used both on MPE and on CPE, as is demonstrated in \cref{fig:attr:host-slave:a}.
They the same apply to function templates, as is shown in \cref{fig:attr:host-slave:b}, and lambda expressions, in \cref{fig:attr:host-slave:c}.
Besides, they also apply to C++ classes to set a default target for their member functions, so that their automatically generated constructors, destructors and assign operators can follow the available target of their belonging class.
Due to the One-Definition Rule of C++, such automatically generated functions are defined at their first usage instead of class definition, thus cannot be covered with the compiler directives introduced later and we have to add attribute support to the classes.

\begin{figure}[htbp]
    \centering

    \begin{subfigure}[t]{\linewidth}
    \begin{minted}[linenos]{cpp}
__attribute((host)) __attribute((slave)) int identity(int x) { return x; }
    \end{minted}
    \subcaption{Example for a function available for both MPE and CPE.}
    \label{fig:attr:host-slave:a}
    \vspace{1em}
    \end{subfigure}

    \begin{subfigure}[t]{\linewidth}
    \begin{minted}[linenos]{cpp}
[&](int x) __attribute((slave)) { return x; }
    \end{minted}
    \subcaption{Example for a CPE lambda expression.}
    \label{fig:attr:host-slave:b}
    \vspace{1em}
    \end{subfigure}

    \begin{subfigure}[t]{\linewidth}
    \begin{minted}[linenos]{cpp}
template <typename T>
__attribute((slave)) T add(T x, T y) { return x + y; }
    \end{minted}
    \subcaption{Example for a CPE function template.}
    \label{fig:attr:host-slave:c}
    \end{subfigure}

    \caption{Example for using \texttt{host} and \texttt{slave} attributes.}
    \label{fig:template}
\end{figure}

\subsubsection{\texttt{infer} attribute}
This attribute primarily targets the usage of user-defined functions (UDFs), in the form of callable template parameters.
Consider the algorithms in C++ Standard Template Library, e.g. \texttt{std::transform}, it is neutral to target by itself, but at the same time limited by the UDF passed to it.
If the UDF is \texttt{slave}, we can only use the \texttt{std::transform} on CPEs, and vice versa.
Thus, function templates like \texttt{std::transform} should determine their instances' available target separately, according to what functions they are calling.
We thus introduce the \texttt{infer} mark, which indicates that the available architecture of the marked function should be the union of all function it calls.
If it calls no function, it is inferred as both \texttt{host} and \texttt{slave}.
We further allow cascading inference to support more generic usages.
The \texttt{infer} attribute is not compatible with other extension attributes.

\subsubsection{\texttt{kernel} attribute}
This attribute specifies a function to be CPE entry points, namely kernel functions.
Calls to such a function has special semantics different from native C++ function calls: they are translated into CPE kernel launches, which starts all the CPEs to run the specified function with the parameters copied to each CPE.
The \texttt{kernel} attribute is not compatible with other extension attributes.

\subsection{Compiler Directives}

To better leverage existing libraries, including C/C++ standard library and ShenWei specific libraries, we introduce a set of compiler directives to specify the available architecture for functions in a sequence of code.
While the global default is \texttt{host}, the directives can change the behavior in part of the code.
Specifically, they include two types of operations:

\begin{itemize}
    \item \texttt{\#pragma swuc push <host|slave|infer>} for setting the default target available for all functions, and save previous default to a stack;
    \item \texttt{\#pragma swuc pop} for recovering the previous default (pop the stack).
\end{itemize}

It is then possible to use STL algorithms through including the header as below:

\begin{minted}[linenos]{cpp}
#pragma swuc push infer
#include <algorithm>
#pragma swuc pop
\end{minted}

\section{Implementation}

To minimize the engineering effort and reuse existing compiler, we implement SWUC as a compiler plugin of the vendor-provided compiler.
The plugin intercepts at middle-end, at which point high-level language features in C++ including template and lambda expression are all lowered into plain intermediate representation.
It helps us to provide natural support for those features without much more effort.
We describe the compilation pipeline as below.

\begin{enumerate}
    \item Separately compile for the MPE and CPE target, each with one invocation of the backend compiler with SWUC plugin loaded:
    \begin{itemize}
        \item for \texttt{host} and \texttt{slave} marked functions, if not available on the currently compiled target, the function body is discarded and not subjected to code generation;
        \item for \texttt{infer} marked functions, we first determine the actual available target(s) through propagating on the call graph, and then filter it out selectively follow the above rule;
        \item for \texttt{kernel} functions, perform differently when generating code for MPE and CPE:
        \begin{enumerate}
            \item for MPE: replace the function body with a wrapper function that performs the kernel launch;
            \item for CPE: additionally generate a CPE-side kernel wrapper function, which unpacks and copies the kernel launch parameters to the CPE stack and call the originally programmed function.
        \end{enumerate}
    \end{itemize}
    \item Link the two machine code object into one, as the compiled artifact of SWUC.
    \item When linking the executable, link necessary runtime library for SWUC into it.
\end{enumerate}

\section{Future Work}

On ShenWei many-core architecture, the MPEs and CPEs do not share the same instruction set as well as vector extension, thus the hardware-related intrinsics are not compatible.
SWUC is not handling the difference currently, and we plan to add support for the intrinsics in the future.
We are also working with multiple groups on different applications to push the adoption of SWUC in real applications, gathering feedback and refine the design.

\section{Conclusion}

ShenWei Universal C/C++ is a language extension for C/C++ to better support heterogeneous programming on ShenWei many-core processors.
With a minimum engineering effort, SWUC manages to unify the programming experience on MPE and CPE cores, simplifying the development of many-core accelerated applications on ShenWei.

\bibliographystyle{ACM-Reference-Format}
\bibliography{main}


\end{document}